\documentstyle[12pt]{article}

\begin{document}

To investigate the question of surface brightness conservation
during (weak and strong) lensing, by using an isothermal sphere
as example,
let $b$ be the the impact
parameter of a ray in the presence of the clump, and $b_0$
that of a ray from the same source point in the absence of the
clump.  Then evidently
 \begin{equation}
b_0=b-L\psi(b),
 \end{equation}
where we defined 
 \begin{equation}
L=\frac{x_l(x_s-x_l)}{(1+z_l)x_s}.
 \end{equation}

Now consider a small element of the source, corresponding to
an element of area of the lensing plane defined by radial and
transverse intervals $db_0$ and $b_0\,d\varphi$.  Its area is
$dA_0=b_0\,db_0\,d\varphi$.  When the clump is in place, the
rays from this element pass through another element with area
$dA=b\,db\,d\varphi$.  The (areal) magnfication of this
element due to lensing is by definition
 \begin{equation}
X=\left|\frac{dA}{dA_0}\right|=
\left|\frac{b\,db}{b_0\,db_0}\right|.
 \end{equation}
The modulus signs are needed of
course because the ratio can be negative when images are
reversed.

Thus the reciprocal magnification is
 \begin{equation}
X^{-1}=\left|\frac{b_0\,db_0}{b\,db}\right|,
 \end{equation}
and from (1) we then find explicitly
 \begin{equation}
X^{-1}=\left|1-L\left(\frac{\psi}{b}+
\frac{d\psi}{db}\right)+L^2\frac{\psi\,d\psi}{b\,db}\right|.
 \end{equation}
Of course, if the lensing is weak, then $X\approx1+2\eta$, with
 \begin{equation}
2\eta=L\left(\frac{\psi}{b}+\frac{d\psi}{db}\right).
 \end{equation}

Another way of expressing the definition (3) is in terms of
the flux ratio.  If the source temperature is $T$, the flux
through the area element in the two cases is
 \begin{equation}
df_0=\sigma T^4\frac{(1+z_l)^2}{x_l^2}dA_0,\qquad
df=\sigma T^4\frac{(1+z_l)^2}{x_l^2}dA,
 \end{equation}
so we could write
 \begin{equation}
X=\frac{df}{df_0}.
 \end{equation}
So long as we are talking about an infinitesimal element, there
is clearly no possibility of violating the conservation of
surface brightness, nor any distinction between flux distance
and area distance.
Is there any reason to think that there could be a violation
for larger areas?

Let's examine the case of a spherically symmetric lens of
total mass $M$ directly in front of a large circular source. 
Suppose that without the lens the source covers a circle of
radius $B_0$ on the lensing plane, where $B_0$ is larger than
the radius $R$ of the lensing mass.  When the lens is present,
the source will then appear expanded to the radius $B$, where
 \begin{equation}
B_0=B-L\psi(B)=B-\frac{4GML}{B}.
 \end{equation}
Without the lens, the total flux from the source is
 \begin{equation}
F_0=\sigma T^4\pi B_0^2\frac{(1+z_l)^2}{x_l^2}.
 \end{equation}
With the lens, assuming surface brightness is conserved, it is
 \begin{equation}
F=\sigma T^4\pi B^2\frac{(1+z_l)^2}{x_l^2}.
 \end{equation}
If $B$ is large (compared to $\sqrt{4GML}$), then the increase
in flux is
 \begin{equation}
F-F_0=8\pi\sigma T^4GML\frac{(1+z_l)^2}{x_l^2},
 \end{equation}
depending only on the total mass of the lens.

What we have to show, to demonstrate consistency, is that this extra
flux is precisely what we should expect from the magnification
of the central parts of the image --- including strong lensing
at the center.
To be more
precise, if we take each small element of area $dA_0$ of the
\emph{unlensed} surface, multiply it by the appropriate
magnification factor $X$ given by (5) and then integrate to
add up these contributions, the result should be precisely the
total area of the surface including the effect of lensing.  In
our case, this means we have to verify that
 \begin{equation}
\int_0^{B_0}2\pi b_0\,db_0\,X_{\rm tot}(b_0)=\pi B^2.
 \end{equation}
For weakly lensed regions, $X_{\rm tot}$ is just the
magnification $X$, but for those regions that are strongly
lensed it is the \emph{sum} of the magnifications of all (in
this case two) images. 

If the density profile is known, we can compute the scattering
angle $\psi$ as a function of $b$.  For our present purposes,
all we really need to know is the behaviour of $\psi(b)$ for
very large and very small values of $b$.  We have already
noted that for large $b$,  
 \begin{equation}
\psi(b)=\frac{4GM}{b},\qquad{\rm for\ }b>R.
 \end{equation}
In the opposite limit, let us assume that at small $b$, as in
the case of a singular isothermal sphere, $\psi$ tends to a
constant:
 \begin{equation}
\psi(b)\approx\psi_0,\qquad{\rm for\ small\ }b.
 \end{equation}

We now need to separate out the region of strong lensing.
The caustic which defines the boundary
of the strong-lensing region is the circle of points that are
images of the central spot (where the magnification is
actually infinite).  The limiting radius $r_c$ is defined by
the condition $b_0=0$, or, by (1),
 \begin{equation}
r_c=L\psi(r_c).
 \end{equation}
For simplicity, let us assume that $r_c$ is sufficiently small
that $\psi$ has already reached its limiting
value $\psi_0$,
as in (15).  Then obviously
 \begin{equation}
r_c=L\psi_0.
 \end{equation}

For $b>r_c$ but close to it, we then have
 \begin{equation}
b_0=b-L\psi_0=b-r_c.
 \end{equation}
When $b$ becomes slightly less than $r_c$, the right hand side
of (18) becomes negative.  What this means is that in fact
 \begin{equation}
b_0=r_c-b,
 \end{equation}
but the source point is on the other side of the center to the
image point, with values of $\varphi$ differing by $\pi$;
images here are reversed.  Thus for every $b_0<r_c$ there are
\emph{two} image circles, with radii
 \begin{equation}
b=r_c\pm b_0.
 \end{equation}
The central disc, of radius $b_0=r_c$, is imaged twice, once
onto the annulus between $b=r_c$ and $b=2r_c$, and once,
reversed, onto the disc $b<r_c$.  In this central region, it's
clear that $|db|=|db_0|$, so the magnifications of the two
images are 
 \begin{equation}
X_\pm=\frac{b}{b_0}=\frac{r_c}{b_0}\pm1.
 \end{equation}
Evidently therefore the total magnification of both images is
 \begin{equation}
X_{\rm tot}=X_++X_-=\frac{2r_c}{b_0}.
 \end{equation}
Hence the contribution of this region to the integral (13) is
 \begin{equation}
\int_0^{r_c}2\pi b_0X_{\rm tot}(b_0)\,db_0=4\pi r_c^2,
 \end{equation}
as we should expect.

For the weakly lensed region, $b>r_c$, we have to evaluate the
integral
 \begin{equation}
I=\int_{r_c}^{B_0}2\pi b_0X(b_0)\,db_0.
 \end{equation}
This is essentially trivial, because $X$ by its definition
is the Jacobian required to change variable from $b_0$ to
$b$.  Thus
 \begin{equation}
I=\int_{2r_c}^{B}2\pi b\,db=\pi B^2-4\pi r_c^2.
 \end{equation}
Adding in the contribution (23) from the central region, this
shows that (13) is indeed satisfied.

\end{document}